\RequirePackage{ifpdf}
\ifpdf 
\documentclass[pdftex]{sigma}
\else
\documentclass{sigma}
\fi

\begin{document}
\allowdisplaybreaks

\renewcommand{\PaperNumber}{074}

\FirstPageHeading

\ShortArticleName{Two- and Three-Particle Correlations in
$q,\!p$-Bose Gas Model}

\ArticleName{Combined Analysis of Two- and Three-Particle\\
Correlations in $\boldsymbol{q,\!p}$-Bose Gas Model}

\Author{Alexandre M. GAVRILIK} 
\AuthorNameForHeading{A.M. Gavrilik}

\Address{N.N. Bogolyubov Institute for Theoretical Physics,
                             Kyiv, Ukraine}
\Email{\href{mailto:omgavr@bitp.kiev.ua}{omgavr@bitp.kiev.ua}}

\ArticleDates{Received December 29, 2005, in f\/inal form October
28, 2006; Published online November 07, 2006}

\Abstract{$q$-deformed oscillators and the $q$-Bose gas model
enable ef\/fective description of the observed non-Bose type
behavior of the intercept (``strength'') $\lambda^{(2)}\equiv
C^{(2)}(K,K)-1$ of two-particle correlation function
$C^{(2)}(p_1,p_2)$ of identical pions produced in heavy-ion
collisions.
  Three- and $n$-particle correlation functions of pions (or kaons)
encode more information on the nature of the emitting sources in
such experiments.  And so, the $q$-Bose gas model was further
developed: the intercepts of $n$-th order correlators of
$q$-bosons and the $n$-particle correlation intercepts within the
$q,\!p$-Bose gas model have been obtained, the result useful for
quantum optics, too. Here we present the combined analysis of two-
and three-pion correlation intercepts for the $q$-Bose gas model
and its $q,\!p$-extension, and confront with empirical data (from
CERN SPS and STAR/RHIC) on pion correlations. Similar to explicit
dependence of $\lambda^{(2)}$ on mean momenta of particles (pions,
kaons) found earlier, here we explore the peculiar behavior,
versus mean momentum, of the 3-particle correlation intercept
$\lambda^{(3)}(K)$.
  The whole approach implies complete chaoticity of sources, unlike
other joint descriptions of two- and three-pion correlations using
two phenomenological parameters (e.g., core-halo fraction plus
partial coherence of sources).}


\Classification{81R50; 81V99; 82B99}

\section{Introduction}

If, instead of treating particles as point-like structure-less
objects, one attempts to take into account either nonzero proper
volume or composite nature of particles, then it is natural to
modify or deform
\cite{Avancini,Perkins} the standard commutation relations. For
these and many other reasons, so-called $q$-deformed oscillators
play important role in modern physics, and various quantum or
$q$-deformed algebras show their ef\/f\/iciency in diverse
problems of quantum physics, see e.g.
\cite{Chang,Chang1,Chang2,q-opt,Gavr1-GI,Gavr1-GI1,Chaichian,
Chaichian1,Gavr2,Cabib}. In quantum optics, $q$-deformed
oscillators and $q$-bosons enable more adequate modeling of the
essentially nonlinear phenomena
\cite{q-opt}. It is worth to mention the usage of both the quantum
counterpart $U_q(su(n))$ of the Lie algebras of f\/lavor groups
$SU(n)$, and the algebras of $q$-deformed oscillators, in the
context of phenomenology of hadron properties
\cite{Gavr1-GI,Gavr1-GI1,Chaichian,Chaichian1,Gavr2,Cabib}.

As it was demonstrated recently, the approach f\/irst proposed
in~\cite{AGI-1,AGI-2} based on some set of $q$-deformed
oscillators along with the related model of ideal gas of
$q$-bosons is quite successful~\cite{AGP} if one  attempts to
ef\/fectively describe the observed, in experiments on
relativistic heavy-ion collisions, non-Bose type properties of the
intercept (``strength'') $\lambda^{(2)}\equiv C^{(2)}(p,p)-1$  of
two-particle correlation function $C^{(2)}(p_1,p_2)$ of the pions
emitted and registered in such experiments. Note that some other
aspects related to this approach were developed
in~\cite{Padula,Padula1}.

Possible reasons for the intercept $\lambda^{(2)}$ to attain (not
the expected Bose value $\lambda^{(2)}=1$, but) the values lesser
than $1$ and thus the reasons to use such $q$-deformed structures
as $q$-oscillators and the related model of ideal gas of
$q$-bosons for an ef\/fective description, include:

\begin{enumerate}
\itemsep=0pt \item[(a)] f\/inite proper volume of particles
\cite{Avancini};

\item[(b)] substructure of particles
\cite{Avancini,Perkins};

\item[(c)] memory ef\/fects;

\item[(d)] ef\/fects from long-lived resonances (e.g., mimicked in
the core-halo picture);

\item[(e)] possible existence of non-chaotic (partially coherent)
components of emitting sources;

\item[(f)] non-Gaussian (ef\/fects of the) sources;

\item[(g)] particle-particle, or particle-medium interactions
(making pion gas non-ideal);

\item[(h)] f\/ireball is a short-lived, highly non-equilibrium and
complicated system.
\end{enumerate}
Clearly, items in this list may be interrelated, say, (a)--(b);
(b)--(c); (e)--(f)--(h).

Three-particle correlation functions $C^{(3)}(p_1,p_2,p_3)$ of
identical pions (or kaons) are as well important, as those carry
additional information
\cite{Heinz,Heinz1} on the space-time geometry and dynamics of the
emitting sources; for the analysis of data, certain combination
\cite{Heinz,Heinz1} of two- and three-pion correlation functions
is usually exploited in which the ef\/fects of long-lived
resonances cancel out.

Further extension of main points of the approach based on $q$-Bose
gas model has been done in        \cite{AdGa}: (i) the intercepts
of $n$-particle correlation functions have been obtained within
$q$-Bose gas model in its two -- Arik--Coon (AC) and
Biedenharn--Macfarlane (BM) -- main versions; (ii) closed
expressions for the intercepts of $n$-particle correlation
functions have been derived using two-parameter ($q,\!p$-)
generalization of the deformed Bose gas model.
 The reason to employ  the $q,\!p$-deformed Bose gas model is two-fold:
f\/irst, the general formulae contain the AC and BM versions as
particular cases; second, the $q,\!p$-Bose gas model provides a
tools to take into account, in a unif\/ied way, any two
independent reasons (to use $q$-deformation) from the above list.

Remark that the results in
\cite{AdGa} are of general value: they can be utilized on an equal
footing both in the domain of quantum optics if $n$-particle
distributions and correlations are important, and in the analysis
of multi-boson (-pion or -kaon) correlations in heavy ion
collisions.

Our goal is to give, within the $q$-Bose gas model and its
two-parameter extended $q,\!p$-Bose gas model, a unif\/ied
analysis of two- and three-particle correlation functions
intercepts using explicit formulas for $C^{(2)}(p,p)$,
$C^{(3)}(p,p,p)$          \cite{AdGa} and a special combination
$r^{(3)}(p,p,p)$                     \cite{Heinz,Heinz1} that
involves both $C^{(2)}$ and $C^{(3)}$. Also, we pay some attention
to a comparison of  analytical results with the existing data on
pion correlations from CERN SPS or STAR/RHIC experiments.

We emphasize a nontrivial shape of dependence on the mean momenta
of pions (or kaons), not only for $C^{(2)}(p,p)$ found earlier but
also for $C^{(3)}(p,p,p)$ and $r^{(3)}(p,p,p)$ studied here: the
dependence turns out to be peculiar for both low and large mean
momenta. The large momentum asymptotics is very special as it is
determined by the deformation parameters $q,\!p$ only.

The whole our treatment assumes {\it complete chaoticity} of
sources, in contrast to other approaches
\cite{Csorgo,Csorgo1,Csorgo2,Csorgo3} to simultaneous description
of two- and three-pion correlations in terms of two
phenomenological parameters where, in the core-halo picture, one
of the parameters ref\/lects partial non-chaoticity or coherence
and the other means a measure of core fraction.

\section[$q$-oscillators and $q,\!p$-oscillators]{$\boldsymbol{q}$-oscillators and $\boldsymbol{q,\!p}$-oscillators}

First, let us recall the well-known {\sl $q$-deformed oscillators
of the Arik--Coon (AC)} type      \cite{AC,AC1,AC2},
\begin{gather}\label{eq1}
 a_i a_j^\dagger - q^{\delta_{ij}} a_j^\dagger a_i=\delta_{ij}  ,
   \qquad
[a_i,a_j]=[a^\dagger_i,a^\dagger_j]=0 ,
\\
[{\cal N}_i,a_j]=-\delta_{ij} a_j ,      \qquad [{\cal
N}_i,a^\dagger_j]=\delta_{ij} a^\dagger_j  , \qquad [{\cal N}_i,
{\cal N}_j]=0 ,\nonumber
\end{gather}
viewed as a set of independent modes. Here and below, $-1\le q\le
1 $. Basis state vectors  $|n_1,\ldots,n_i,\ldots\rangle$ are
constructed from vacuum state $|0,0,\ldots\rangle $ as usual, and
the operators $a^\dagger_i$ act with matrix elements
$\langle\ldots,n_i+1,\ldots | a^\dagger_i|\ldots,n_i,\ldots\rangle
= \sqrt{\lfloor n_i+1\rfloor } $ where the ``basic numbers''
$\lfloor r\rfloor\equiv {(1-q^r)}/{(1-q)}$ are used. The
$q$-bracket $\lfloor A\rfloor$ for an operator $A$ means formal
series. As the $q$-{parameter} $q\to 1$, the $\lfloor r\rfloor$
resp. $\lfloor A\rfloor$ goes back to $r$ resp. $A$. The operators
$a^\dagger_i ,$ $a_i$ are mutual conjugates if $-1\le q\le 1$.
Note that the equality $a^\dagger_i a_i={\cal N}_i$ holds only at
$q=1$, while for $q\ne 1$ the $a^\dagger_i a_i$ {\it depends on
the number operator} ${\cal N}_i$ {\it nonlinearly}:
\begin{equation} \label{eq2}
a^\dagger_i a_i=\lfloor{\cal N}_i\rfloor .
\end{equation}
The system of the {\sl two-parameter extended or $q,\!p$-deformed
oscillators} is def\/ined as
\cite{Chakra}
\begin{equation}\label{eq3}
A_i A_j^\dagger - q^{\delta_{ij}} A_j^\dagger A_i=
                          \delta_{ij} p^{N_i^{(qp)}} , \qquad
 A_i A_j^\dagger - p^{\delta_{ij}} A_j^\dagger A_i=
                           \delta_{ij} q^{N_i^{(qp)}} ,
\end{equation}
along with the relations $ [N^{(qp)},A]=-A$,
$[N^{(qp)},A^\dagger]= A^\dagger$.
  For the $q,\!p$-deformed oscillators,
\begin{equation}\label{eq4}
A^\dagger A=[\![N^{(qp)}]\!]_{qp} ,
                   \qquad
[\![X]\!]_{qp} \equiv  (q^{X}-p^{X})/(q-p)   .
\end{equation}
At $p=1$ the AC-type $q$-bosons are recovered.

On the other hand, putting $p=q^{-1}$ in \eqref{eq3} yields the
other important for our treatment case, {\sl the $q$-deformed
oscillators of Biedenharn--Macfarlane (BM)} type~\cite{BM,BM1}
 with the main relation
\begin{equation}\label{eq5}
b_i b_j^\dagger-q^{\delta_{ij}} b_j^\dagger
b_i=\delta_{ij}q^{-N_j} .
\end{equation}
The ($q$-)deformed Fock space is constructed likewise, but now,
instead of basic numbers, we use the other $q$-bracket and
``$q$-numbers'' (compare this with the brackets used in
\eqref{eq2} and \eqref{eq4}):
\begin{equation}\label{eq6}
b^\dagger_i b_i=[N_i]_q  ,   \qquad [r]_q \equiv
(q^r-q^{-r})/(q-q^{-1})  .
\end{equation}
{} The equality $b^\dagger_i b_i = N_i$ holds only if $q=1$. For
consistency of conjugation we put
\begin{equation}\label{eq7}
 q=\exp (i \theta) ,   \qquad   0 \le \theta < \pi .
\end{equation}

\section[$q$-deformed and $q,\!p$-deformed momentum distributions]{$\boldsymbol{q}$-deformed and $\boldsymbol{q,\!p}$-deformed momentum distributions}
\label{sec3}


Dynamical multi-particle (multi-pion, -kaon, -photon, \dots)
system will be viewed as an ideal gas of $q$- or $q,\!p$-bosons
whose statistical properties are described by evaluating the
thermal averages
\begin{equation}\label{eq8}
\langle A \rangle= \frac{{\rm Sp} (A e^{-\beta H} ) }{  {\rm Sp}(
e^{-\beta H} ) } , \qquad H=\sum_i{\omega_i {\cal N}_i} , \qquad
\omega_i=\sqrt{m^2+{\boldsymbol k}_i^2} ,
\end{equation}
where summation runs over dif\/ferent modes, $\beta=1/T$, and the
Boltzmann constant is set $k=1$. In the Hamiltonian in
\eqref{eq8}, ${\cal N}_i$ denotes the number operator of the
respective version: the AC-type, BM-type, or $q,\!p$-type. The
$3$-momenta of particles are assumed to be discrete-valued.

The $q$-distribution for AC-type $q$-bosons ($ -1\le q\le 1 $) is
$ \langle a_i^\dagger a_j\rangle=\delta_{ij}\langle a_i^\dagger
a_i\rangle =\delta_{ij}/(e^{\beta\omega_i}-q) , $ reducing to the
usual Bose--Einstein distribution if $q\to 1$. From now on, all
the formulas will correspond to the mono-mode case (coinciding
modes), and we shall omit the indices.

At $q=0$ or $q=-1$, the distribution function $\langle a^\dagger
a\rangle=(e^{\beta\omega}-q)^{-1} $ yields the familiar classical
Boltzmann or the Fermi--Dirac cases (the latter only formally:
dif\/fering modes of $q$-bosons at $q=-1$ are commuting, unlike
genuine fermions whose non-coinciding modes anticommute).

The one-particle momentum distribution function for the AC-type
$q$-bosons, along with one-particle distributions for the BM-type
$q$-bosons and for the general $q,\!p$-Bose gas case, are placed
in the following Table \ref{table1} (f\/irst column). The second
column of this same Table contains the two-particle mono-mode
momentum distributions, correspondingly, for the three mentioned
cases of deformed Bose gas.
  The results concerning one-particle distributions
are known from the papers~\cite{Vokos,Vokos1,Vokos2} whereas those
on two-particle distributions from~\cite{AGI-1,AGI-2,Kibler}.

\begin{table}[t]\centering
\caption{One-particle and two-particle distributions of the ($q$-
or $qp$-) deformed bosons.}\label{table1}

\vspace{1mm}

\begin{tabular}{|c|l|l|}
\hline
&&\\[-3mm]
     {Case}  & \multicolumn{1}{|c|}{1-particle distribution} &  \multicolumn{1}{|c|}{2-particle distribution}   \\[1mm]
\hline
{} & {} & {} \\[-3mm]
AC-case  & $\langle a^\dagger a
\rangle=\frac{1}{e^{\beta\omega}-q}$  & $\langle  {a^\dagger}^2
a^2 \rangle = \frac{ (1+q) }
{(e^{\beta\omega}-q)(e^{\beta\omega}-q^2) }$   \\[1.5mm]
\hline
{} & {} & {} \\[-3mm]
BM-case  & $\langle b^\dagger b \rangle=\frac{e^{\beta\omega}-1}
{e^{2\beta\omega}- (q+q^{-1}) e^{\beta\omega}+1} $           &
$\langle {b^\dagger}^2 b^2 \rangle = \frac{ (q+q^{-1})
}
{(e^{\beta\omega}-q^2) 
(e^{\beta\omega}-q^{-2})}$   \\[1.5mm]
\hline
{} & {} & {} \\[-3mm]
$q,\!p$-bosons  & $\langle  A^\dagger A \rangle = \frac{
(e^{\beta\omega}-1) }
       { (e^{\beta\omega}-p) (e^{\beta\omega}-q) }$           &
       $\langle {A^\dagger}^2  A^2 \rangle = \frac{ (p+q)
(e^{\beta\omega}-1) }
{(e^{\beta\omega}-q^2)(e^{\beta\omega}- p q)(e^{\beta\omega}-p^2)}$    \\[2mm]
\hline
\end{tabular}
\end{table}

Few remarks are in order. (i) The $q$-distribution functions for
the AC case (f\/irst row) are real for real deformation parameter
$q$. The $q$-distribution functions for the BM case (second row)
are real not only with real deformation parameter, but also for
$q=\exp(i\theta)$, in which case we have:
$q+q^{-1}=[2]_q=2\cos\theta$, and
$q^2+q^{-2}=[2]_{q^2}=2\cos(2\theta)$. The $q,\!p$-distributions,
see third row, are real if both $q$ and $p$ are real, or $|q|=1$
and $p=q^{-1}$, or $p=\bar{q}$. (ii) Each of the two one-particle
$q$-deformed distributions (AC-type or BM-type $f_q({\boldsymbol
k})\equiv \langle b^\dagger b\rangle ({\boldsymbol k})$) is such
that at $q\ne 1$ the corresponding curve is intermediate between
the familiar Bose--Einstein and Boltzmann curves. (iii)
Generalized $q,\!p$-deformed one- and two-particle distribution
functions given in third row, reduce to the corresponding
distributions of the $q$-{bosons} of AC-type in the f\/irst row if
$p=1$ (BM type in the second row if $p=q^{-1}$). (iv) Here and
below, all the two-parameter $q,\!p$-expressions possess the
interchange symmetry under $q\leftrightarrow p$.

\section[$n$-particle distributions and correlations
           of deformed bosons]{$\boldsymbol{n}$-particle distributions and correlations
           of deformed bosons}         \label{sec4}


The most general result, derived in
\cite{AdGa} and based on the $q,\!p$-oscillators, for the
$n$-particle distribution functions of the gas of $q,\!p$-bosons
\begin{equation}\label{eq9}
\langle { A^\dagger}^n   A^n \rangle = \frac{   [\![n]\!]_{qp}!\
(e^{\beta\omega}-1)} {\prod\limits_{r=o}^n ( e^{\beta\omega} - q^r
p^{n-r} ) }  ,
 \qquad [\![m]\!]_{qp}!=[\![1]\!]_{qp}[\![2]\!]_{qp}
\cdots [\![m-1]\!]_{qp}[\![m]\!]_{qp}  ,
\end{equation}
involves the $q,\!p$-bracket def\/ined in \eqref{eq4}. Let us
remark that this general formula for the $n$-th order, or
$n$-particle, mono-mode momentum distribution functions for the
ideal gas of $q,\!p$-bosons follows from combining the following
two relations
\begin{equation}\label{eq10}
\langle A^{\dag n} A^{n} \rangle =
\frac{[\![n]\!]_{qp}!}{\prod_{r=0}^{n-1} (e^{\beta\omega}-p^r
q^{n-r}) } {\langle p^{nN} \rangle }             \qquad   {\rm
and}  \qquad {\langle p^{nN}\rangle}
=\frac{e^{\beta\omega}-1}{e^{\beta\omega}-p^n} .
\end{equation}
Below, we are interested in the $q,\!p$-deformed intercept
$\lambda^{(n)}_{q,p}  \equiv - 1 +
 \frac{{\langle A^{\dagger n} A^{n} \rangle}}{{\langle
A^{\dagger} A \rangle}^{n}}  $ of $n$-particle correlation
function. The corresponding formula constitutes main result in
\cite{AdGa} and reads:
\begin{equation}     \label{eq11}
 \lambda^{(n)}_{q,p}
= [\![n]\!]_{qp}! \frac{(e^{\beta\omega}-p)^n
(e^{\beta\omega}-q)^n} {(e^{\beta\omega}-1)^{n-1}
\prod\limits_{k=0}^n(e^{\beta\omega}-q^{n-k}p^k)} - 1 .
\end{equation}
Consider the asymptotics $\beta\omega\to\infty$ (for large momenta
or, with f\/ixed momentum, for low temperature) of the intercepts
$\lambda^{(n)}_{q,p}$:
\begin{equation}\label{eq12}
\lambda^{(n), \, {\rm asym.} }_{q,p} = - 1 + [\![n]\!]_{qp}! = - 1
+ \prod^{n-1}_{k=1}\biggl(\sum^k_{r=0}q^r p^{k-r}\biggl)  .
\end{equation}
It is worth noting that for each case: the $q$-bosons of AC-type,
of BM-type, and the $q\!p$-bosons, the asymptotics of $n$-th order
intercept is given by the corresponding deformed extension of
$n$-factorial (the intercept of pure Bose--Einstein $n$-particle
correlator is given by the usual $n!$).

To specialize these results for particular case of AC-type
$q$-bosons we set $p=1$. The $n$-particle distribution function
and the $n$-th order intercept $\lambda^{(n)} \equiv \frac{
\langle a^{\dagger n} a^n\rangle } { \langle a^\dagger a\rangle^n
} - 1 $ take the form
\begin{equation}\label{eq13}
\langle  {a^\dagger}^n a^n \rangle = \frac{\lfloor n\rfloor!
}{\prod\limits_{r=1}^n (e^{\beta\omega}-q^r) } ,        \qquad
\lambda^{(n)}_{\rm {AC}}= - 1 + \frac{ \lfloor n\rfloor! \
(e^{\beta\omega}-q)^{n-1} } {
\prod\limits^n_{r=2}(e^{\beta\omega}-q^r) } ,
\end{equation}
where $\lfloor m\rfloor \equiv
\frac{1-q^m}{1-q}=1+q+q^2+\dots+q^{m-1}$. At
$\beta\omega\to\infty$ the result involves only $q$-parameter:
\begin{gather*}
\lambda^{(n), \, {\rm asym.}}_{\rm {AC}} = -1 + \lfloor n\rfloor!
= -1+
\prod^{n-1}_{k=1}\biggl(\sum^k_{r=0}q^r\biggr)\\
\phantom{\lambda^{(n), \, {\rm asym.}}_{\rm {AC}}}{} =
(1+q)(1+q+q^2)\cdots(1+q+\cdots+q^{n-1})-1 .
\end{gather*}
This constitutes a principal consequence of the approach. It would
be very nice to verify this, using the data for pions and kaons
drawn from the experiments on relativistic nuclear collisions.

The expressions for the BM case that are parallel to \eqref{eq13}
follow from the general formulas~\eqref{eq9}, \eqref{eq11},
\eqref{eq12} if we put $p=q^{-1}$.


{\bf Intercepts of two- and three-particle correlation functions,
              their asymptotics.}
Let us consider the 2nd and 3rd order correlation intercepts. In
Table 2 we present the intercepts of two-particle correlations for
the $q$-bosons of AC-type and BM-type, as well as for the general
case of $q,\!p$-bosons. Recall that at $p=q^{-1}$,
$[\![2]\!]_{qp}\equiv p+q $ reduces to
$[2]_q=q+q^{-1}=2\cos\theta$.

\begin{table}[t]\centering
\caption{Intercept of two-particle correlations of deformed bosons
              and its asymptotics.}\label{table2}

\vspace{1mm}

\begin{tabular}{|c|l|l|}
\hline
&&\\[-3mm]
 {\rm Case}  & \multicolumn{1}{|c|}{Intercept $\lambda^{(2)}\equiv
                             \frac{ \langle a^{\dagger 2} a^2\rangle }
 {\langle a^\dagger a\rangle^2 }-1 $} &
            \multicolumn{1}{|c|}{Asymptotics of $\lambda^{(2)}$ at $\beta\omega\to\infty$}  \\[2mm]
\hline
{} & {} & {} \\[-3mm]
AC  &  $\lambda^{(2)}_{\rm AC}= \frac{(1+q)(e^{\beta\omega}-q)}
{e^{\beta\omega}-q^2}-1=
q\frac{e^{\beta\omega}-1}{e^{\beta\omega}-q^2}$   &
 $\lambda^{(2),\, {\rm asym.}}_{\rm AC}= q $   \\[1.5mm]
\hline
{} & {} & {} \\[-3mm]
BM  & $\lambda^{(2)}_{\rm BM}= \frac{2\cos\theta
(e^{2\beta\omega}-2\cos\theta\ e^{\beta\omega}+1)^2}
{(e^{\beta\omega}-1)^2(e^{2\beta\omega} -2\cos(2\theta)
e^{\beta\omega}+1)}-1$
 &  $\lambda^{(2),\, {\rm asym.}}_{\rm BM} = 2\cos\theta - 1 $   \\[1.5mm]
\hline
{} & {} & {} \\[-3mm]
$q,\!p$  &  $\lambda^{(2)}_{q,p} = \frac{(p+q)
(e^{\beta\omega}-p)^2 (e^{\beta\omega}-q)^2}
{(e^{\beta\omega}-1)(e^{\beta\omega}-q^2)(e^{\beta\omega}- p q)
(e^{\beta\omega}-p^2)} -1$ & $\lambda^{(2),\, {\rm asym.}}_{q,p} = (p+q)-1$  \\[1.5mm]
\hline
\end{tabular}
\end{table}

In the non-deformed limit $q\to 1$ (or $\theta\to 0$) the value
$\lambda_{\rm BE}=1$ known for Bose--Einstein statistics is
correctly reproduced from the formulas in the f\/irst two rows of
Table~\ref{table2}. This obviously corresponds to the
Bose--Einstein distribution recovered from the $q$-Bose one at
$q\to 1$.
Now consider three-particle correlations. The intercepts for all
the three versions of deformed Bose gas along with their
asymptotics are presented in Table~\ref{table3} as three rows
correspondingly.
 We end with two remarks valid for both the Table~\ref{table2} and Table~\ref{table3}.
First, the formulas for AC- or BM-type follow from those of the
$q,\!p$-Bose gas (given in the third row) if we set $p=1$ or
$p=q^{-1}$. Second, as demonstrated by the last column, the large
$\beta\omega$ asymptotics of each of the three versions of
$\lambda^{(3)}$ is given by a very simple dependence on the
deformation parameter(s) only.

\begin{table}[t]\centering
\caption{Intercepts of three-particle correlations of deformed
bosons
              and their asymptotics.}\label{table3}
\vspace{1mm}

\begin{tabular}{|c|l|l|}
\hline
&&\\[-3mm]
{\rm Case} &
    \multicolumn{1}{|c|}{Intercept $\lambda^{(3)}\equiv \frac{ \langle a^{\dagger 3} a^3\rangle }
            { \langle a^\dagger a\rangle^3 }-1$}  &
            \multicolumn{1}{|c|}{Asymptotics of $\lambda^{(3)}$ at $\beta\omega\to\infty$}  \\[2mm]
\hline
{} & {} & {} \\[-3mm]
AC  & $\lambda^{(3)}_{\rm AC} =
\frac{(1+q)(1+q+q^2)(e^{\beta\omega}-q)^2}
                           {(e^{\beta\omega}-q^2)(e^{\beta\omega}-q^3)}-1$
&  $\lambda^{(3),\, {\rm asym.}}_{\rm AC}= (1+q)(1+q+q^2) - 1 \!\!$     \\[1.5mm]
\hline
{} & {} & {} \\[-3mm]
BM  &   $\lambda^{(3)}_{\rm BM} =
\frac{2\cos\!\theta(4\cos^2\theta-1)(e^{2\beta\omega}
-2\cos\!\theta\ e^{\beta\omega} +1)^2}
     {(e^{\beta\omega}-1)^2 (e^{2\beta\omega}
-2\cos(3\theta) e^{\beta\omega}+1)}-1 $ &
$\lambda^{(3),\, {\rm asym.}}_{\rm BM}= 2\cos\theta (4\cos^2\theta-1) -1\!\!$ \\[1.5mm]
\hline
{} & {} & {} \\[-3mm]
$q,\!p$  & $\lambda^{(3)}_{q,p}\!= \!\frac{ (p+q)(p^2+pq+q^2)
(e^{\beta\omega}-p)^3 (e^{\beta\omega}-q)^3}
{(e^{\beta\omega}-1)^2 (e^{\beta\omega}-p^3) (e^{\beta\omega}-p^2
q)(e^{\beta\omega}-pq^2) (e^{\beta\omega}-q^3)}\!-\!1\!\!$
&
$\lambda^{(3),\, {\rm asym.}}_{q,p}\!=\!(p+q)(p^2+pq+q^2)\!-\!1\!\! $    \\[1.5mm]
\hline
\end{tabular}
\end{table}

\section{Comparison with data on two- and three-pion correlations}
                              \label{sec5}

\begin{figure}[t]
\centering {\includegraphics[angle=0,
width=0.6\textwidth]{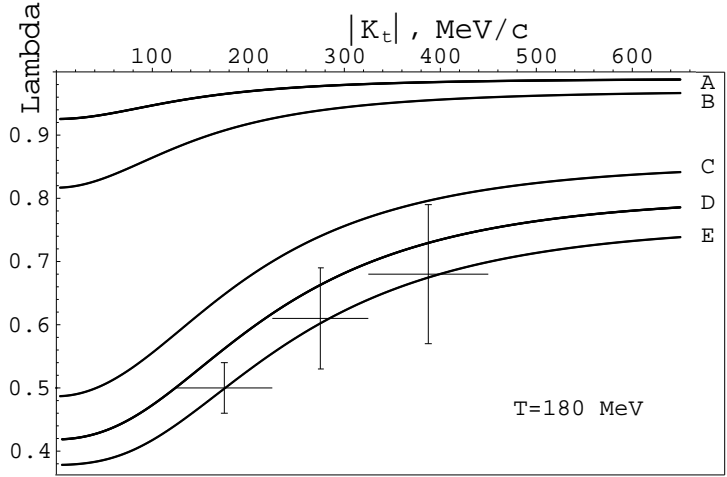}} \caption{Intercept
$\lambda^{(2)}_{\rm BM}$ versus pions' transverse momentum $K_{\rm
t}$, at $T=180$ MeV and $q=\exp(i\theta)$: A)~$\theta = 6^\circ$;
B)~$\theta = 10^\circ$;  C)~$\theta = 22^\circ$; D)~$\theta =
25.7^\circ$, i.e.\ $2 \theta_c$; E)~$\theta = 28.5^\circ$. The
$\pi^-\pi^-$ data are taken from~\cite{Adams,Adams1}.}\label{fig1}
\end{figure}

The explicit dependence, via $\beta\omega=\frac1T\sqrt{m^2+K_{\rm
t}^2}$, of the intercept $\lambda^{(2)}_{\rm BM}$ on transverse
mean momentum is shown in Fig.~\ref{fig1} where the empirical
values on {\it negative} pions~\cite{Adams,Adams1}, for three
momentum bins (horizontal bars), are shown as the
crosses\footnote{Vertical bars characterize the experimental
uncertainty. See more detailed comments in
\cite{Gavr2,AGP}.}. At $T\!=\!180$ MeV, these three values agree
perfectly with the curve E for which $q = \exp(i\theta)$,\ $\theta
= 28.5^\circ $. Slightly higher temperature $T=205$ MeV  gives
nice agreement of empirical values with the curve  of
$\theta=\frac{\pi}{7}\simeq 25.7^\circ $, twice the Cabibbo
angle\footnote{In~\cite{Gavr2,Cabib}, detailed arguments can be
found on the possible link of the $q$-deformation parameter with
the Cabibbo angle, in case of the quantum algebra $su_q(3)$
applied to static properties of the octet hadrons. Remark also
that the gauge interaction eigenstates of quarks are the
Cabibbo-mixed superpositions of their mass
eigenstates~\cite{Cheng}.}. We may conjecture that, this way, {\em
the hot pions in their correlations show} some ``memory'' of their
origin as quark-antiquark bound states, see the items (b)--(c) in
Introduction. Possible relation of the Cabibbo (quark mixing)
angle to the value $q$ of deformation which measures deviation of
the $q$-Bose gas picture from pure Bose one, looks quite
promising. However, more empirical data for two-pion (-kaon)
correlations with more bins for diverse momenta in experiments on
relativistic nuclear collisions are needed in order to conf\/irm
within this model the general trend and the key feature that for
$\beta\omega\to \infty$ (large momenta, or low temperatures) the
intercept does saturate with a value given by deformation
parameter(s) only.

  Equally good comparison of the data for two-particle
correlations of {\it positive} pions~\cite{Adams,Adams1} with the
AC-type of $q$-Bose gas model at the optimal $q=0.63$ was
achieved~\cite{AGP}, and general observation was made that the
AC-type $q$-Bose gas model with {\em real} $q$ is well suited for
treating $\pi^+\pi^+$ correlations, while the BM version with
$q=e^{i\theta}$ serves better for $\pi^-\pi^-$ correlations.

\begin{figure}[t]
\centering {\includegraphics[angle=-90,
width=0.55\textwidth]{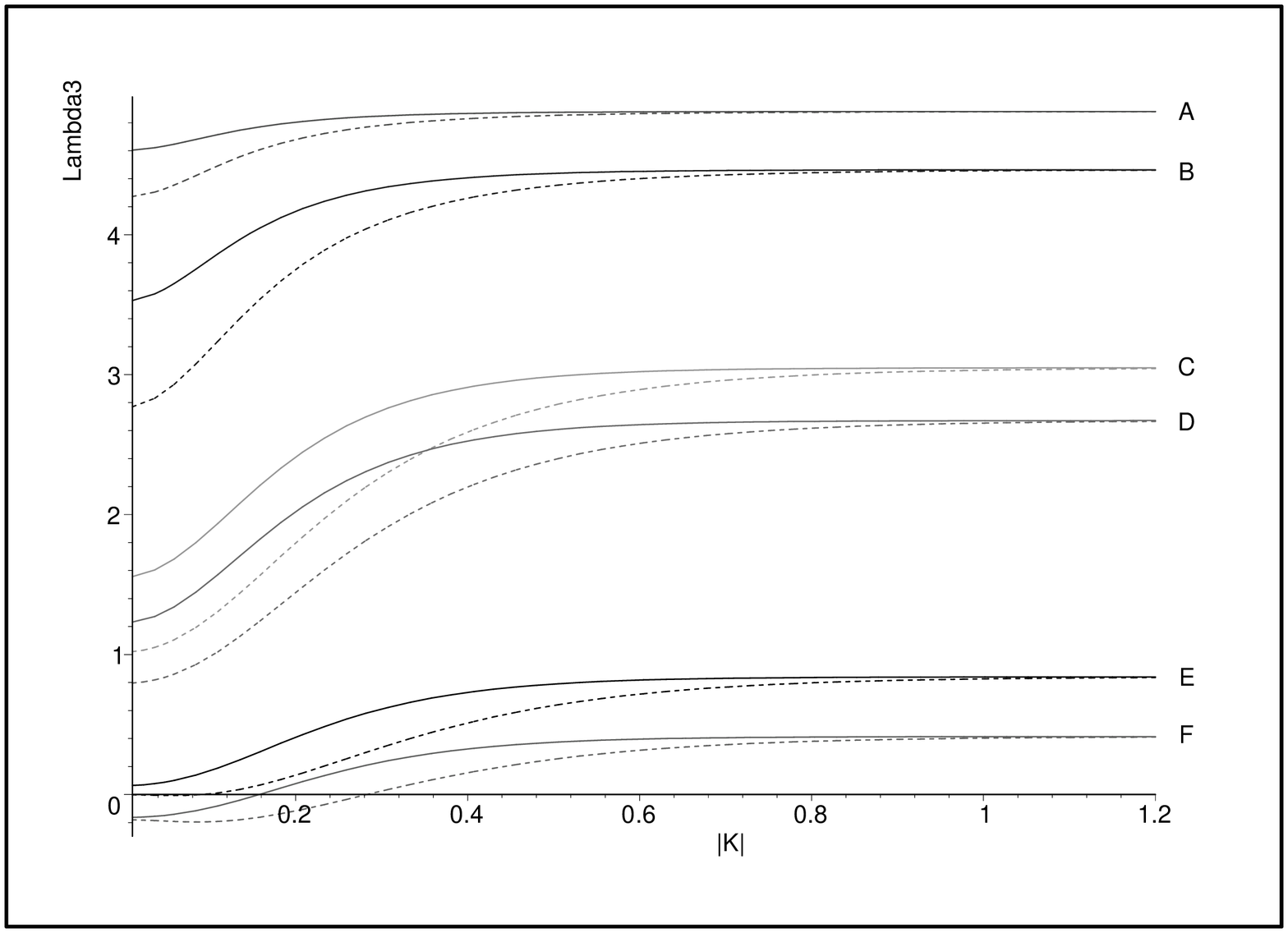}}
\caption{Intercept $\lambda_{\rm BM}^{(3)}$ \ versus pions' 
mean momentum $|{\bf K}|$, GeV/c. The temperature is 120 MeV (180
MeV) for solid (dashed) curves. In each pair the two curves have
common asymptotics given, see Table 3, by the value of $\theta$,
from the top down: $\frac{\pi}{30}$~(A);  $\frac{\pi}{14}$~(B);
$\frac{\pi}{7}$~(C); $\frac{28.5\pi}{180}$~(D);
$\frac{9.26\pi}{40}$~(E);  $\frac{\pi}{4}$~(F).}\label{fig2}
\end{figure}

In Fig.~\ref{fig2} we plot the dependence on particles' 
mean momentum for the 3rd order correlation intercept
$\lambda_{\rm BM}^{(3)}$ of the BM version of $q$-Bose gas.
Observe the asymptotic saturation with constant values f\/ixed by
$q$ (or $\theta$), as Table \ref{table3} prescribes. Also, a
peculiar behavior of the intercept $\lambda_{\rm BM}^{(3)}$ is
evident: at low momenta it acquires zero or even negative values
for some curves, see ``E'', ``F''. However, these two curves imply
large angles $\theta$ ({\em large deviations from pure Bose value}
$\theta=0$), the cases hardly realizable in view of the items
(a)-(h) of Introduction. Curves similar to those in
Fig.~\ref{fig2} can be presented for the 3rd order intercept
$\lambda_{\rm AC}^{(3)}$ of the AC-type of $q$-Bose gas model.

\begin{figure}[t]
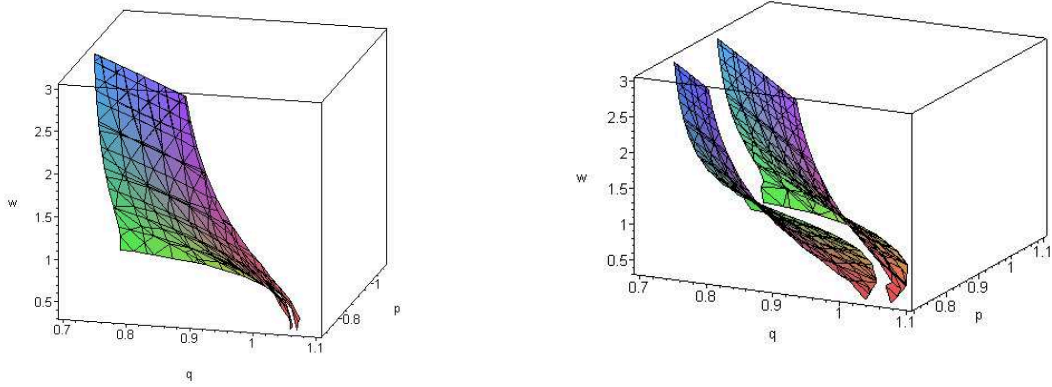
   
\centering {\begin{minipage}{7.2cm}
\includegraphics[angle=0, width=7.2cm]{gavrilik-fig3i}
 \end{minipage}
 \quad
 \begin{minipage}{8.2cm}
 \includegraphics[angle=0, width=8.2cm]{gavrilik-fig3ai}
 \hspace{-14mm}
\end{minipage}}
\vspace{-5mm} \caption{{\em Left}: The two ``central'' surfaces as
implicit functions $w=w(q,\!p)$: one stems from equating
$\lambda^{(2)}_{p,q}(w)$ from Table \ref{table2} to the {\sl
central} value $\lambda^{(2), {\rm exp.}}=0.57$ in \eqref{eq14},
the other stems from equating $\lambda^{(3)}_{p,q}(w)$ from
Table~\ref{table3} to the {\sl central} value $\lambda^{(3), {\rm
exp.}}=1.92$ in \eqref{eq15}. Notice mutual closeness of these
``central'' surfaces. {\em Right}: similar to the left panel, but
now $\lambda^{(2)}_{p,q}(w)$ is equated to  $\lambda^{(2), {\rm
exp.}}_+=0.57+0.04$ in \eqref{eq14}
and $\lambda^{(3)}_{p,q}(w)$  
to $\lambda^{(3), {\rm exp.}}_-=1.92-0.49$ in
\eqref{eq15}.}\label{fig3}
\end{figure}

Now compare the explicit form of the intercepts
$\lambda^{(2)}_{p,q}$ and $\lambda^{(3)}_{p,q}$ of two- and
three-particle correlations of $p,\!q$-bosons from Tables
\ref{table2}, \ref{table3} with the available data on the 2nd and
3rd order correlations of pions
\cite{Adams,Adams1,Bearden,Bearden1} from the experiments on
relativistic nuclear collisions, namely
\cite{Bearden,Bearden1}
\begin{gather}
   \lambda^{(2), {\rm exp.}}|_{\rm neg.pions}=0.57\pm 0.04  ,     \label{eq14}
\\
    \lambda^{(3), {\rm exp.}}|_{\rm neg.pions}=1.92\pm 0.49 .\label{eq15}
\end{gather}
Equating the expression for say $\lambda^{(2)}_{p,q}(w)$ from
Table~\ref{table2} (with $w\equiv\beta\omega$) to some f\/ixed
value, in particular those in \eqref{eq14}, \eqref{eq15}, leads to
an implicit function $w=w(q,\!p)$ whose image is 2-surface.
Fig.~\ref{fig3} (left) shows the two surfaces obtained from
equating $\lambda^{(2)}_{p,q}$ resp.\ $\lambda^{(3)}_{p,q}$ to the
{\it central values} of the data $\lambda^{(2),{\rm exp}}$ resp.
$\lambda^{(3),{\rm exp}}$ in \eqref{eq14}, \eqref{eq15}. As seen,
these ``central'' surfaces are very close to each other, for both
low and large values of $w$ (recall, $w\equiv\beta\omega=
T^{-1}\sqrt{m^2+K_{\rm t}^2}$); only in their lower right corner
the surfaces {\sl are seen} as disjoint. 
For comparison, in Fig.~\ref{fig3} (right) we show the result of
equating $\lambda^{(2)}_{p,q}(w)$ resp.\ $\lambda^{(3)}_{p,q}(w)$
to the upper value $\lambda^{(2), {\rm exp.}}_+=0.57 + 0.04$
resp.\ lower value $\lambda^{(3), {\rm exp.}}_-=1.92 - 0.49$ in
\eqref{eq14}, \eqref{eq15}. Here the resulting two surfaces are
clearly distant. Likewise, using the values $\lambda^{(2), {\rm
exp.}}_-=0.57\!-\!0.04$, $\lambda^{(3), {\rm exp.}}_+=1.92+0.49$
in \eqref{eq14}, \eqref{eq15}, we get yet another two surfaces
which we do not exhibit. 
In total, due to \eqref{eq14}, \eqref{eq15} we have six surfaces:
the three (named ``$\lambda^{(2)}$-triple'') obtained by equating
$\lambda^{(2)}_{p,q}(w)$ to each of the 3 values $\lambda^{(2),
{\rm exp.}}$ in~\eqref{eq14}, and the other three (named
``$\lambda^{(3)}$-triple'') got by equating
$\lambda^{(3)}_{p,q}(w)$ to each value in~\eqref{eq15}.

\begin{figure}[t]
\centerline{\begin{minipage}{7cm}
\includegraphics[angle=0, width=7cm]{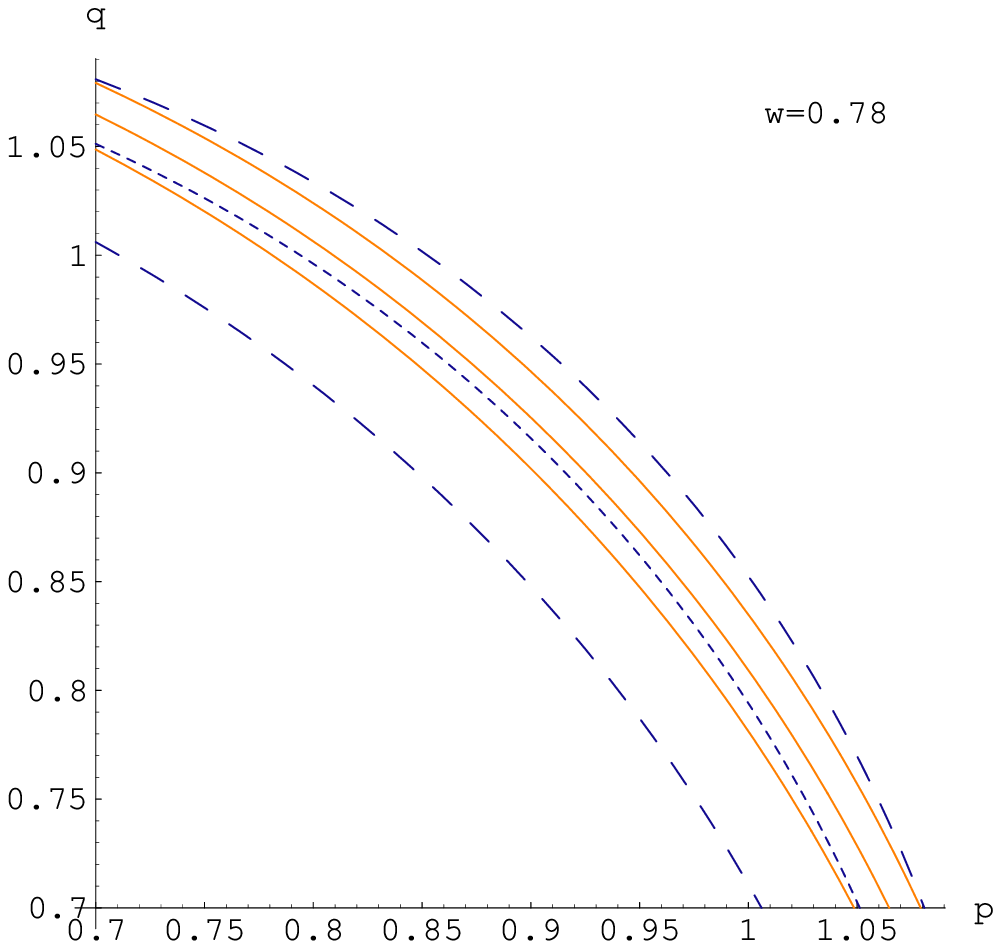}
 \end{minipage}
 \qquad
 \begin{minipage}{7cm}
 \includegraphics[angle=0, width=7cm]{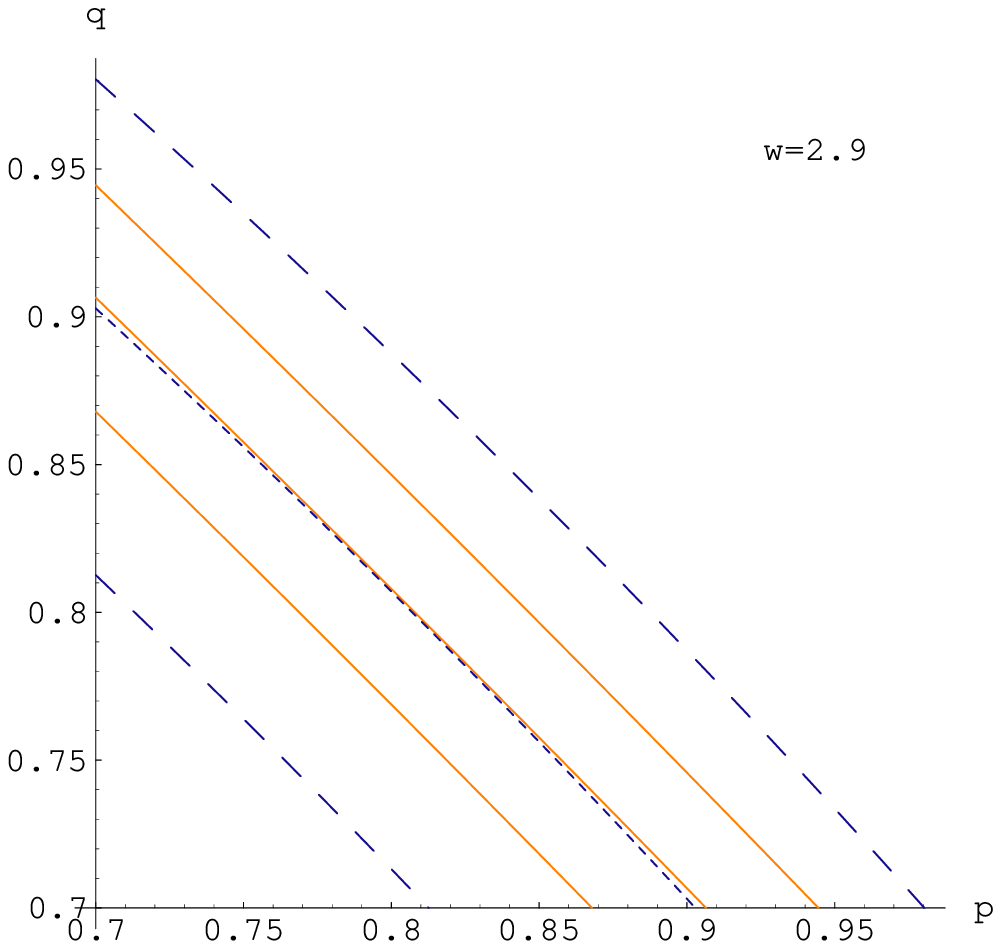}
\end{minipage}}

\caption{{\em Left}: Slice of all the six surfaces (two pairs in
Fig.~\ref{fig3} and the remaining 2 surfaces, see main text) by
putting $w=0.78$. Solid (dashed) lines stem from cutting the
$\lambda^{(2)}$-triple (the $\lambda^{(3)}$-triple) of surfaces.
{\em Right}: same as left panel, but now $w=2.9$ is put. Notice
full overlapping of $\lambda^{(2)}$-strip and
$\lambda^{(3)}$-strip.}\label{fig4}
\end{figure}

Cutting the both triples of surfaces by a horizontal plane $w=w_0$
with some f\/ixed $w_0$ yields, in this plane, six curves given as
six implicit functions $q=q(p)$. In Fig.~\ref{fig4}, we exhibit
the result of slicing these two triples by $w = 0.78$ (left
panel), and by $w = 2.9$ (right panel)\footnote{For the {\em gas
of pions} whose mass $m_{\pi}=139.57$ MeV, setting the temperature
and the momentum as $T=180$ MeV and $K_{\rm t}=10$ MeV yields
$w\approx 0.78$, while $T=180$ MeV and $K_{\rm t}=500$ MeV yields
$w\approx 2.9$.}.
In each panel we observe that (the strip formed by)
$\lambda^{(3)}$-triple of curves fully covers (the strip of)
$\lambda^{(2)}$-triple.
Such full covering dif\/fers from the situation
in~\cite{Csorgo,Csorgo1,Csorgo2,Csorgo3} where only the imaging of
data with~2$\sigma$ (or 3$\sigma$) uncertainty gave some
overlapping. Also unlike~\cite{Csorgo,Csorgo1,Csorgo2,Csorgo3}, we
present {\em explicit momentum dependence} for $\lambda^{(2)}$, $
\lambda^{(3)}$. Besides, our approach assumes complete chaoticity
of particle sources.

Let us make few remarks concerning Figs.~\ref{fig3} and
\ref{fig4}.

1) The two very near ``central'' surfaces of Fig.~\ref{fig3}
(left) ref\/lect themselves in Fig.~\ref{fig4} as follows: the
result of their slicing by $w=0.78$ is seen in the left panel as
the two very close lines, one solid and one dashed, which connect
the points given roughly as (1.05, 0) and (0, 1.05), while the
result of their slicing by $w=2.9$ is seen in the right panel as
the two almost straight, almost coinciding solid and dashed lines
given roughly by the equation $q=0.9 - p$.
On the other hand, the two distant surfaces of Fig.~\ref{fig3}
(right) yield after their slicing the lowest dashed and the
uppermost solid lines in each panel of Fig.~\ref{fig4}.

2) In both left and right panels of Fig.~\ref{fig4}, the entire
strip formed by the three solid (i.e., $\lambda^{(2)}$-) curves,
lies completely within the strip formed by the three dashed
$\lambda^{(3)}$-curves. The larger width of the latter strip is
due to better accuracy of data in \eqref{eq14} as compared with
that in \eqref{eq15}.

3) Limiting ourselves to one-parameter cases, from
Figs.~\ref{fig3}, \ref{fig4} we deduce: beside the AC-type ($p=1$)
and BM-type ($p=q^{-1}$) $q$-oscillators def\/ined in \eqref{eq1}
and \eqref{eq5} and their versions of $q$-Bose gas, there is yet
another distinguished case of $q$-oscillator and so yet another
version of $q$-Bose gas model equally well suited for explaining
the experimental data: this is the
$q$-oscillator\footnote{Following~\cite{TD,TD1} we call it the
Tamm--Dancof\/f (or TD) deformed oscillator.}
which is contained in the relations \eqref{eq3}, \eqref{eq4} at
$p=q$ and which also leads, by applying $p=q$, to the
corresponding formulas for distributions, intercepts etc. for this
version of $q$-Bose gas.

4) It is easily seen from Fig.~\ref{fig4} that the TD-type of
$q$-Bose gas model constructed on the base of the TD-type
$q$-oscillator (see footnote~4) is more preferable than the AC version
for description of the data like \eqref{eq14} and \eqref{eq15}:
f\/irst, it needs narrower range of values of the $q$-parameter in
order to cover the data \eqref{eq14}, \eqref{eq15} including the
uncertainties; second, with just these data the use of AC version
is problematic for the large momenta region because of $p=1$, see
right panel of Fig.~\ref{fig4}, while on the contrary the TD
version corresponding to $p=q$ serves equally well for both small
($w=0.78$) and large ($w=2.9$) values of momenta.

{\bf Unif\/ied analysis of $\lambda^{(2)}$ and $\lambda^{(3)}$
versus empirical data using special combination.} For a unif\/ied
analysis of data on the two- and three-particle correlations,
there exists~\cite{Heinz,Heinz1} a~very convenient special
combination designed in terms of correlators, namely
\[
\hspace{-6mm}    r^{(3)}(p_1,p_2,p_3) \equiv \frac12
\frac{C^{(3)}(p_1,p_2,p_3)-C^{(2)}(p_1,p_2)
-C^{(2)}(p_2,p_3)-C^{(2)}(p_3,p_1)+2}
{\sqrt{\left(C^{(2)}(p_1,p_2)-1\right)\left(C^{(2)}(p_2,p_3)-1\right)
\left(C^{(2)}(p_3,p_1)-1\right)}},  
\]
which with the restriction $p_1=p_2=p_3=K$ turns into the simple
expression
\begin{equation} \label{eq16}
r^{(3)}_j(K) \equiv r^{(3)}_j(K,K,K)=\frac12
\frac{\lambda_j^{(3)}(K)-3\lambda_j^{(2)}(K) }
{\bigl(\lambda_j^{(2)}(K)\bigr)^{3/2} }
\end{equation}
composed of just the intercepts. Here $\lambda^{(2)}(K)
=C^{(2)}(K,K)-1$, $\lambda^{(3)}(K)=C^{(3)}(K,K,K)-1$, and the
subscript $j$ denotes the particular type (AC, BM, TD, or
$q,\!p$-) of the deformed Bose gas. The relevant formulas for the
intercepts involved are to be taken from Tables~\ref{table2},
\ref{table3}. The convenience of \eqref{eq16} from the viewpoint
of comparison with data is two-fold:  f\/irst, the design 
of~$r^{(3)}$ is such that the contribution to it from long-lived
resonances cancels out~\cite{Heinz,Heinz1}; second, the~$r^{(3)}$
carries an additional meaning being a (cosine of) special
phase~\cite{phase}.

\begin{figure}[t]
\centerline{\includegraphics[angle=-90.0,
width=8cm]{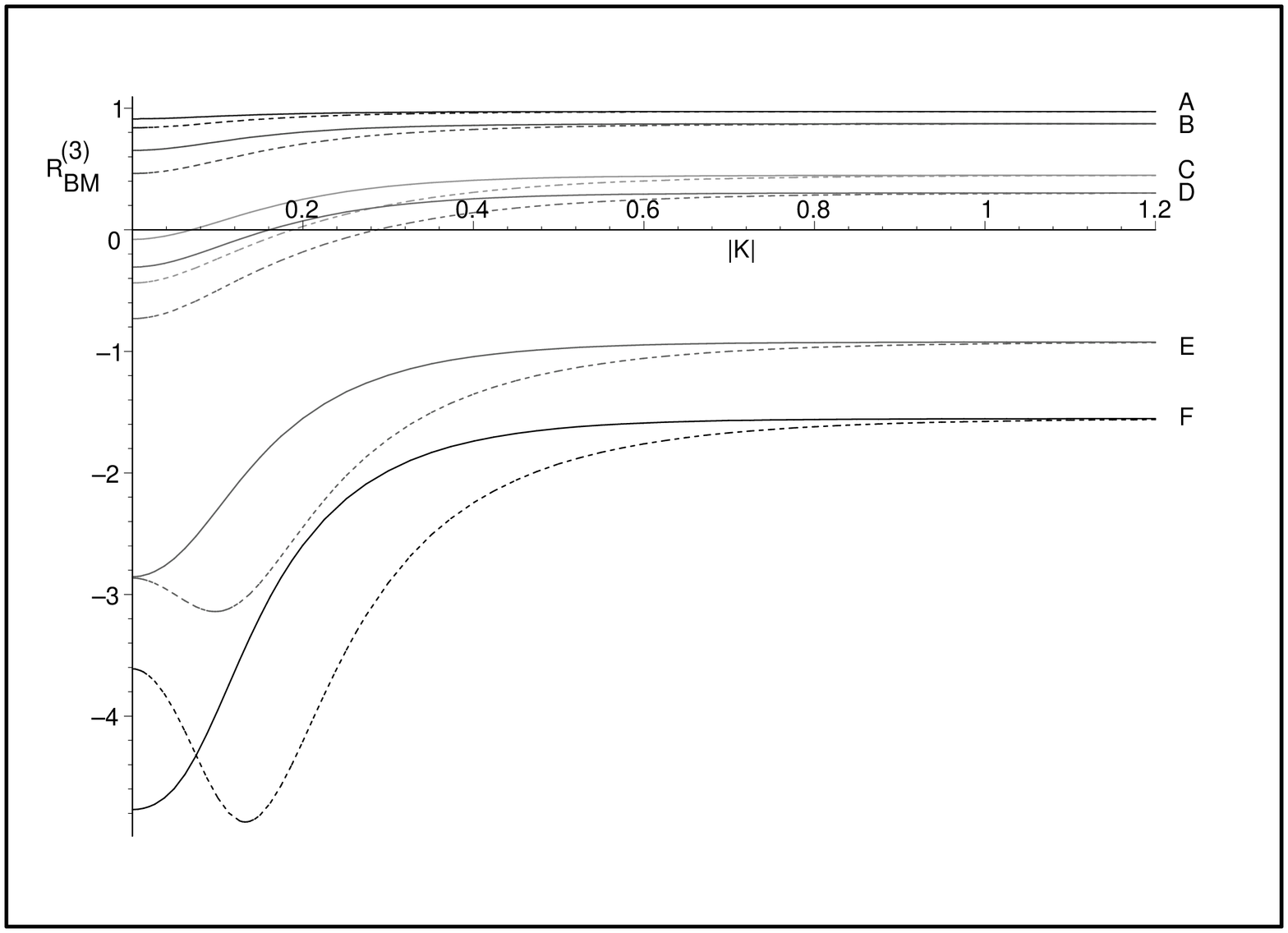}} \caption{Plot of $r^{(3)}_{\rm BM}({\bf
K})$ from (16) with the intercepts $\lambda^{(2)}_{\rm BM}$,
$\lambda^{(3)}_{\rm BM}$, versus pions' mean momentum $|{\bf K}|$,
GeV/c. For solid (dashed) lines $T=120$ MeV ($T=180$ MeV). In each
pair, the two curves have common asymptotics determined by its
$\theta$ value: $\frac{\pi}{30}$~(A), $\frac{\pi}{14}$~(B),
$\frac{\pi}{7}$~(C), $\frac{28.5\pi}{180}$~(D),
$\frac{9.26\pi}{40}$~(E), $\frac{\pi}{4}$~(F).}\label{fig5}
\end{figure}

In Fig.~\ref{fig5}, taking as example the one-parameter BM version
of $q$-Bose gas model, we show main features of the momentum
dependence of $r_{\rm BM}^{(3)}({\bf K})$. Again the behavior is
rather peculiar as it shows nontrivial shape in the small $|{\bf
K}|$ region, asymptotic saturation with a constant determined by
$q$ (or $\theta$) at large $|{\bf K}|$, and the property that the
value of temperature $T$ is very important.
  Of course, it will be very interesting to check the validity
of such a behavior by confronting with relevant empirical data. We
only note that the presently available data on
the~$r^{(3)}$~values~\cite{Adams,Adams1} are controversial (the
set of values ranges from zero to unity) and insuf\/f\/icient,
usually not indicating to which momenta bins they correspond.

      \section{Conclusions and outlook}

We have demonstrated the ef\/f\/iciency of the generalized
$q,\!p$-deformed Bose gas model in combined description of the
intercepts of two- and three-particle correlations, especially in
view of the non-Bose type properties of the 2nd and 3rd order
correlations of pions (and kaons) observed in the experiments on
relativistic collisions of heavy nuclei. The key advantages of the
model are:
 (i)~{\em it provides the explicit dependence} on particle's mean 
momentum of the intercepts $\lambda^{(2)}$, $\lambda^{(3)}$ as
well as of their combination $r^{(3)}$ given in (16); (ii)~{\em it
possesses the asymptotical property} that at
$\beta\omega\to\infty$ each of these quantities becomes
independent of particles' mass, momentum and temperature and takes
a very simple form determined by the deformation parameters
$q,\!p$ only. The property (ii) is important as it enables, using
the empirical data on $\lambda^{(2)}$ and $\lambda^{(3)}$ {\sl for
sufficiently large transverse momenta}, to determine with high
accuracy the values of the deformation parameters $q, p$ which, in
accord with our paradigm, should characterize the nontrivial and
special system under study, see the items (a)--(h) of
Introduction. With so f\/ixed deformation parameters and due to
(i), one should {\sl use the low transverse momentum data} for
$\lambda^{(2)}$ and $\lambda^{(3)}$ in order to f\/irmly f\/ix the
value of temperature.

A remark on the one-parameter limiting cases of the $q,\!p$-Bose
gas model. From the Figs.~\ref{fig3},~\ref{fig4} we have deduced
that, beside the most popular BM- and AC-cases of $q$-Bose gas
model, yet another distinguished one-parameter case is of
interest: this is the Tamm--Dancof\/f (TD) version of $q$-deformed
oscillators and $q$-Bose gas model, got from the general
two-parameter $q,\!p$-formulas by putting $p=q$. The TD case
provides as well an appropriate basis for the analysis of data on
two-, three- and possibly multi-particle momentum correlators
(f\/irst of all, correlation intercepts) of pions and kaons from
the experiments like those reported
in~\cite{Adams,Adams1,Bearden,Bearden1}.

We have seen from the Figs.~\ref{fig1}, \ref{fig3}, \ref{fig4},
that the $q,\!p$-Bose gas model, which is an extension of the
$q$-Bose gas model, is in agreement with presently available
experimental data. Concerning the data from
experiments~\cite{Adams,Adams1} especially those on the intercept
of 3-pion correlations and the quantity $r^{(3)}$, it is highly
desirable to obtain not only the values averaged over large range
of transverse mean momenta or transverse mass of identical
particles, but also a more detailed data with numerous momentum
bins, {\sl for both small and large transverse momenta}. A rich
enough set of momentum-attributed values of $\lambda^{(3)}$ and
$r^{(3)}$ will allow to draw more certain conclusions about the
viability of the considered model. With detailed experimental
data, further judgements will be possible about actual physical
meaning (recall the items (a)--(h) in the Introduction) and
adequate values of the deformation parameter(s) $q$ or $q,\!p$.
This includes the special one-parameter case of the BM-type
$q$-Bose gas model with its possible link~\cite{Gavr2,Cabib} of
the $q$-parameter $\ q=\exp(i\theta)$ to the Cabibbo (quark
mixing) angle.

\looseness=1
Of course, more work is needed to further develop the approach
based on the concept of $q,\!p$-Bose gas. Note that in our study
of the implications of the $q,\!p$-Bose gas model and subsequent
analysis of relevant data, we dealt only with the intercept (i.e.\
the strength) of the two-pion as well as three- and multi-pion
correlation functions. Put in another words, we treated only the
mono-mode (single-mode) case which means coinciding momenta of the
correlated particles.
Since within general $q,\!p$-deformation of Bose gas model it is
desirable to have a complete correlation functions with full
momentum dependence including nonzero relative momenta of
particles, clearly the multi-mode case should be elaborated.
Although the modeling of complete correlation functions is highly
non-unique (see e.g.~\cite{non-gauss}) and not too trivial, as a
f\/irst step one may proceed in analogy to the one parameter
$q$-Bose gas model (e.g.\ like in~\cite{Padula}).

\subsection*{Acknowledgements}

The author is grateful to Professor R.~Jagannathan for sending a
copy of his paper. This work is partially supported by the Grant
10.01/015 of the State Foundation of Fundamental Research of
Ukraine.

\LastPageEnding


\begin{thebibliography}{99}
\footnotesize

\bibitem{Avancini}  Avancini S.S., Krein G.,
Many-body problems with composite particles and $q$-Heisenberg
algebras, {\it  J.~Phys.~A: Math. Gen.}, 1995, V.28, 685--691.

\bibitem{Perkins}     Perkins W.A.,     Quasibosons,
{\it Internat. J. Theoret. Phys.}, 2002, V.41, 823--838,
\href{http://arxiv.org/abs/hep-th/0107003}{hep-th/0107003}.

\bibitem{Chang}   Chang Z.,
Quantum group and quantum symmetry, {\it Phys. Rep.}, 1995, V.262,
137--225,
\href{http://arxiv.org/abs/hep-th/9508170}{hep-th/9508170}.

\bibitem{Chang1} Kibler M.R., Introduction to quantum algebras,  \href{http://arxiv.org/abs/hep-th/9409012}{hep-th/9409012}.

\bibitem{Chang2} Mishra A.K., Rajasekaran G.,
Generalized Fock spaces, new forms of quantum statistics and their
algebras, {\it Pramana}, 1995, V.45, 91--139,
\href{http://arxiv.org/abs/hep-th/9605204}{hep-th/9605204}.

\bibitem{q-opt}  Man'ko V.I., Marmo G., Sudarshan E.C.G., Zaccaria F.,
$f$-oscillators and nonlinear coherent states, {\it Phys.
Scripta}, 1997, V.55, 528--541,
\href{http://arxiv.org/abs/quant-ph/9612006}{quant-ph/9612006}.

\bibitem{Gavr1-GI}  Gavrilik A.M.,
$q$-Serre relations in $U_q(u_n)$, $q$-deformed meson mass sum
rules, and Alexander polynomials, {\it J. Phys.~A: Math. Gen.},
1994, V.27, L91--L94.

\bibitem{Gavr1-GI1}
Gavrilik A.M., Iorgov N.Z., Quantum groups as f\/lavor symmetries:
account of nonpolynomial $SU(3)$-breaking ef\/fects in baryon
masses, {\it Ukrain. J. Phys.}, 1998, V.43, 1526--1533,
\href{http://arxiv.org/abs/hep-ph/9807559}{hep-ph/9807559}.

\bibitem{Chaichian}   Chaichian M., Gomez J.F., Kulish P.,
Operator formalism of q deformed dual string model, {\it Phys.
Lett. B}, 1993, V.311, 93--97,
\href{http://arxiv.org/abs/hep-th/9211029}{hep-th/9211029}.

\bibitem{Chaichian1}   Jenkovszky L., Kibler M., Mishchenko A.,
Two-parametric quantum deformed dual amplitude, {\it Modern Phys.
Lett. B}, 1995, V.10, 51--60,
\href{http://arxiv.org/abs/hep-ph/9407071}{hep-ph/9407071}.

\bibitem{Gavr2}  Gavrilik A.M.,
Quantum algebras in phenomenological description of particle
properties, {\it Nucl. Phys. B (Proc. Suppl.)}, 2001, V.102,
298--305,
\href{http://arxiv.org/abs/hep-ph/0103325}{hep-ph/0103325}.

\bibitem{Cabib}
Gavrilik A.M., Quantum groups and Cabibbo mixing, in Proceedings
of Fifth International Confe\-ren\-ce ``Symmetry in Nonlinear
Mathematical Physics'' (June 23--29, 2003, Kyiv), Editors
A.G.~Nikitin, V.M.~Boyko, R.O.~Popovych and I.A.~Yehorchenko, {\it
Proceedings of Institute of Mathematics}, Kyiv, 2004, V.50,
Part~2, 759--766,
\href{http://arxiv.org/abs/hep-ph/0401086}{hep-ph/0401086}.

\bibitem{AGI-1}   Anchishkin D.V., Gavrilik A.M., Iorgov N.Z.,
Two-particle correlations from the $q$-Boson viewpoint, {\it Eur.
Phys. J. A}, 2000, V.7, 229--238,
\href{http://arxiv.org/abs/nucl-th/9906034}{nucl-th/9906034}.

\bibitem{AGI-2}  Anchishkin D.V., Gavrilik A.M., Iorgov N.Z.,
$q$-Boson approach to multiparticle correlations, {\it Modern
Phys. Lett.~A}, 2000, V.15, 1637--1646,
\href{http://arxiv.org/abs/hep-ph/0010019}{hep-ph/0010019}.

\bibitem{AGP}   Anchishkin D.V., Gavrilik A.M., Panitkin S.,
Intercept parameter $\lambda$ of two-pion (-kaon) correlation
functions in the $q$-boson model: character of its
$p_T$-dependence, {\it Ukrain. J. Phys.}, 2004, V.49, 935--939,
\href{http://arxiv.org/abs/hep-ph/0112262}{\mbox{hep-ph/0112262}}.

\bibitem{Padula}   Zhang Q.H., Padula S.S.,
$Q$-boson interferometry and generalized Wigner function,
 {\it Phys. Rev. C}, 2004, V.69, 24907, 11 pages,  \href{http://arxiv.org/abs/nucl-th/0211057}{nucl-th/0211057}.

\bibitem{Padula1}
Gutierrez T., Intensity interferometry with anyons,
 {\it Phys. Rev. A}, 2004, V.69, 063614, 5 pages,   \href{http://arxiv.org/abs/quant-ph/0308046}{\mbox{quant-ph/0308046}}.

\bibitem{Heinz}
Wiedemann U.A., Heinz U., Particle interferometry for relativistic
heavy-ion collisions, {\it Phys. Rept.}, 1999, V.319, 145--230,
\href{http://arxiv.org/abs/nucl-th/9901094}{nucl-th/9901094}.

\bibitem{Heinz1} Heinz U., Zhang Q.H.,
What can we learn from three-pion interferometry? {\it Phys. Rev.
C}1997, V.56, 426--431,
\href{http://arxiv.org/abs/nucl-th/9701023}{nucl-th/9701023}.


\bibitem{AdGa}
Adamska L.V., Gavrilik A.M., Multi-particle correlations in
$qp$-Bose gas model, {\it J. Phys.~A: Math. Gen.}, 2004, V.37,
N~17, 4787--4795,
\href{http://arxiv.org/abs/hep-ph/0312390}{hep-ph/0312390}.

\bibitem{Csorgo}
Csorg\"o T. et al., Partial coherence in the core/halo picture of
Bose--Einstein $n$-particle correlations, {\it Eur. Phys. J. C},
1999, V.9, 275--281,
\href{http://arxiv.org/abs/hep-ph/9812422}{hep-ph/9812422}.

\bibitem{Csorgo1}
Csanad M. for PHENIX Collab., Measurement and analysis of  two-
and three-particle correlations,
\href{http://arxiv.org/abs/nucl-ex/0509042}{\mbox{nucl-ex/0509042}}.

\bibitem{Csorgo2}
Morita K., Muroya S., Nakamura H., Source chaoticity from two- and
three-pion correlations in Au+Au collisions at $\sqrt{s_{NN}}=130$
GeV, \href{http://arxiv.org/abs/nucl-th/0310057}{nucl-th/0310057}.

\bibitem{Csorgo3}
Biyajima M., Kaneyama M., Mizoguchi T., Analyses  of two- and
three-pion Bose--Einstein correlations using Coulomb wave
functions, {\it Phys. Lett. B}, 2004, V.601, 41--50,
\href{http://arxiv.org/abs/nucl-th/0312083}{nucl-th/0312083}.

\bibitem{AC}
Arik M., Coon D.D., Hilbert spaces of analytic functions and
generalized coherent states, {\it J. Math. Phys.},  1976, V.17,
524--527.

\bibitem{AC1}
Fairlie D., Zachos C., Multiparameter associative generalizations
of canonical commutation relations and quantized planes,
 {\it Phys. Lett. B},  1991, V.256, 43--49.

\bibitem{AC2}
Meljanac S., Perica A., Generalized quon statistics,
{\it Modern Phys. Lett.~A}, 1994, V.9, 3293--3300.  %


\bibitem{Chakra}   Chakrabarti A., Jagannathan R.,
A ($p,q$)oscillator realization of two-parameter quantum algebras,
{\it J.~Phys.~A: Math. Gen.}, 1991, V.24, L711--L718.

\bibitem{BM}   Macfarlane A.J.,
On $q$-analogues of the quantum harmonic oscillator and the
quantum group $SU_q(2)$, {\it  J.~Phys.~A: Math. Gen.}, 1989,
V.22, 4581--4585.

\bibitem{BM1}
Biedenharn L.C., The quantum group $SU_q(2)$ and a $q$-analogue of
the boson operators, {\it J.  Phys. A: Math. Gen.}, 1989, V.22,
L873--L878.

\bibitem{Vokos}
Altherr T., Grandou T., Thermal f\/ield theory and inf\/inite
statistics,
 {\it Nucl. Phys. B}, 1993, V.402, 195--216.

\bibitem{Vokos1}
Vokos S., Zachos C., Thermodynamic $q$-distributions that aren't,
{\it Modern Phys. Lett. A}, 1994, V.9, 1--9.

\bibitem{Vokos2}
Lavagno A., Narayana Swamy P.,  Thermostatistics of $q$ deformed
boson gas, {\it Phys. Rev. E}, 2000, V.61, 1218--1226,
\href{http://arxiv.org/abs/quant-ph/9912111}{quant-ph/9912111}.

\bibitem{Kibler}
Daoud M., Kibler M., Statistical mechanics of $qp$-bosons in $D$
dimensions, {\it Phys. Lett. A}, 1995, V.206, 13--17,
\href{http://arxiv.org/abs/quant-ph/9512006}{quant-ph/9512006}.

\bibitem{Adams}
Adler C. et al. [STAR Collab.], Pion interferometry of
$\sqrt{S(NN)}$ = 130-GeV Au+Au collisions at RHIC, {\it Phys. Rev.
Lett.}, 2001, V.87, 082301, 6 pages,
\href{http://arxiv.org/abs/nucl-ex/0107008}{nucl-ex/0107008}.

\bibitem{Adams1}
Adams J. et al. [STAR Collab.], Three-pion HBT correlations in
relativistic heavy-ion collisions from the STAR experiment, {\it
Phys. Rev. Lett.}, 2003, V.91, 262301, 6 pages,
\href{http://arxiv.org/abs/nucl-ex/0306028}{nucl-ex/0306028}.

\bibitem{Bearden}
Aggarwal M.M. et al. [WA98 Collab.], One-, two- and three-particle
distributions from 158$A$ GeV/$c$ central Pb+Pb collisions,
 {\it Phys. Rev. C}, 2003, V.67, 014906, 24 pages, \href{http://arxiv.org/abs/nucl-ex/0210002}{nucl-ex/0210002}.

\bibitem{Bearden1}
Bearden I.G. et al. [NA44 Collab.], One-dimensional and
two-dimensional analysis of 3 pi correlations measured in Pb + Pb
interactions,  {\it Phys. Lett. B}, 2001, V.517, 25--31,
 \href{http://arxiv.org/abs/nucl-ex/0102013}{nucl-ex/0102013}.

\bibitem{Cheng}
Cheng T.-P., Li L.-F., Gauge theory of elementary particle
physics,
 Oxford, Clarendon Press, 1984.

\bibitem{TD}
Odaka K., Kishi T., Kamefuchi S., On quantization of simple
harmonic oscillators, {\it  J. Phys. A: Math. Gen.}, 1991, V.24,
L591--L596.

\bibitem{TD1}
Chaturvedi S., Srinivasan V., Jagannathan R., Tamm--Dancof\/f
deformation of bosonic oscillator algebras, {\it Modern Phys.
Lett. A}, 1993, V.8, 3727--3734.

\bibitem{phase}  Heiselberg H., Vischer A.P.,
The phase in three-pion correlations,
\href{http://arxiv.org/abs/nucl-th/9707036}{nucl-th/9707036}.

\bibitem{non-gauss} Csorgo T., Szerzo A.T.,
Model independent shape analysis of correlations in 1, 2 or 3
dimensions, {\it Phys. Lett. B}, 2000, V.489, 15--23,
\href{http://arxiv.org/abs/hep-ph/0011320}{hep-ph/0011320}.

\end{thebibliography}
\end{document}